\documentclass[aps,twocolumn,groupedaddress,showpacs,floatfix]{revtex4}

\begin{document}
\bibliographystyle{apsrev}

\title{Detecting topological order in a ground state wave function}

\author{Michael Levin}
\author{Xiao-Gang Wen}
\affiliation{Department of Physics, Massachusetts Institute of Technology,
Cambridge, Massachusetts 02139}

\begin{abstract}
A large class of topological orders can be understood and classified using the
string-net condensation picture. These topological orders can be 
characterized by a set of data $(N,d_i,F^{ijk}_{lmn},\del_{ijk})$. We describe 
a way to detect this kind of topological order using only the ground state wave
function. The method involves computing a quantity called the ``topological 
entropy'' which directly measures the quantum dimension $D=\sum_i d^2_i$.
\end{abstract}

\maketitle

\textsl{Introduction}:
Until recently, the only known physical characterizations of topological order
\cite{Wtoprev} involved properties of the Hamiltonian - e.g. quasiparticle
statistics \cite{ASW8422}, ground state degeneracy \cite{HR8529,WNtop}, and 
edge excitations \cite{Wtoprev}. In this paper, we demonstrate that topological
order is manifest not only in these dynamical properties but also in the basic
entanglement of the ground state wave function. We hope that this 
characterization of topological order can be used as a theoretical tool to 
classify trial wave functions - such as resonating dimer wave functions 
\cite{RK8876}, Gutzwiller projected states,
\cite{BA8880,AM8874,WWZcsp,RS9173,Wsrvb} or quantum loop gas wave functions 
\cite{FNS0428}. In addition, it may be useful as a numerical test for 
topological order. Finally, it demonstrates definitively that topological order
is a property of a wave function, not a Hamiltonian. The classification of 
topologically ordered states is nothing but a classification of complex 
functions of thermodynamically large numbers of variables.

\begin{figure}[tb]
\centerline{
\includegraphics[width=2.2in]{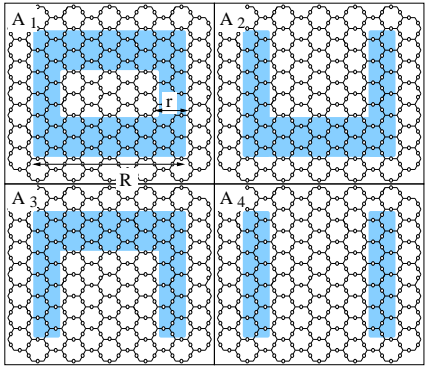}
}
\caption{
One can detect topological order in a state $\Psi$ by computing the von Neumann
entropies $S_1, S_2, S_3, S_4$ of the above four regions, $A_1, A_2, A_3, A_4$,
in the limit of $R,r \rightarrow \infty$.  Here the four regions are drawn in
the case of the honeycomb lattice. The geometry ensures that the number of
sites $n_1,n_2,n_3,n_4$ along the boundaries of the $4$ regions obey the
relation $n_1 - n_2 = n_3 - n_4$. 
}
\label{Z2topent1}
\end{figure}

\textsl{Main Result}:
We focus on the $(2+1)$ dimensional case (though the result can be generalized
to any dimension). Let $\Psi$ be an arbitrary wave function for some two
dimensional lattice model. For any subset $A$ of the lattice, one can compute
the associated quantum entanglement entropy $S_A$.\cite{HIZ0515} The main
result of this paper is that one can determine the  ``quantum dimension'' $D$
of $\Psi$ by computing the entanglement entropy $S_A$ of particular regions $A$
in the plane. Normal states have $D=1$ while topologically ordered states have
$D \neq 1$ (e.g. in the case of topological orders described by discrete gauge 
theories, $D$ is equal to the number of elements in the gauge group). Thus, $D$
provides a way to distinguish topologically ordered states from normal states.

More specifically, consider the four regions $A_1,A_2,A_3,A_4$ drawn in
Fig. \ref{Z2topent1}. Let the corresponding entanglement entropies be 
$S_1,S_2,S_3,S_4$. Consider the linear combination $(S_1-S_2)-(S_3-S_4)$,
computed in the limit of large, thick annuli, $R,r \rightarrow \infty$. The 
main result of this paper is that 
\begin{equation}
\label{quantdim}
(S_{1}-S_{2})-(S_{3}-S_{4}) = -\log(D^2)
\end{equation}
where $D$ is the quantum dimension of the topological field theory associated
with $\Psi$. 

We call the quantity $(S_{1}-S_{2})-(S_{3}-S_{4})$ the ``topological entropy'',
$-S_{\text{top}}$, since it measures the entropy associated with the 
(non-local) topological entanglement in $\Psi$. The above result implies that
$S_\text{top}$ is an universal number associated with each topological phase.
It is invariant under smooth deformations of $\Psi$. 

\textsl{Physical picture}:
The idea behind \Eq{quantdim} is that topologically ordered states contain
nonlocal entanglement. Consider, for example, a spin-$1/2$ model with spins 
located on the links $\v i$ of the honeycomb lattice and with a Hamiltonian 
realizing a $Z_2$ lattice gauge theory. \cite{K032,W7159,K7959} The ground 
state wave function $\Psi$ is known exactly. The easiest way to describe $\Psi$
is in terms of strings. One can think of each spin state as a string state, 
where a $\si_{\v i}^x = -1$ spin corresponds to a link occupied by a string and
a $\si_{\v i}^x = 1$ spin corresponds to an empty link. In this language, 
$\Psi$ is very simple: $\Psi(X) = 1$ for all string states $X$ where the 
strings form closed loops, and $\Psi$ vanishes otherwise. 

\begin{figure}[tb]
\centerline{
\includegraphics[width=1.2in]{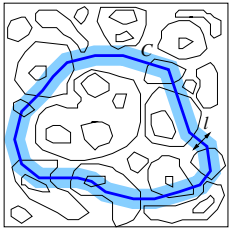}
}
\caption{
The state $\Psi$ contains nonlocal correlations 
originating from the fact that closed strings always cross a 
closed curve $C$ an even number of times. These correlations can be measured by
a string operator $W(C)$ (thin blue curve). For more general states, a 
fattened string operator $W_{\text{fat}}(C)$ (light blue region) is 
necessary.
}
\label{physpict}
\end{figure}

All local correlations $\<\si_{\v i}^x\si_{\v j}^x\>$ vanish for this state.
However, $\Psi$ contains \emph{nonlocal} correlations or entanglement. To see
this, imagine drawing a curve $C$ in the plane (see Fig. \ref{physpict}). There
is a nonlocal correlation between the spins on the links crossing this curve:
$\<W(C)\> = \<\prod_{\v i \in C} \si_{\v i}^x\> = 1$. This correlation
originates from the fact that the number of strings crossing the curve is
always even. Similar correlations exist for more general states
that contain virtual string-breaking fluctuations. In the general case, the 
nonlocal correlations can be captured by ``fattened'' string operators 
$W_{\text{fat}}(C)$ that act on spins within some distance $l$ of $C$ where $l$
is the length scale for string breaking.

To determine whether a state is topologically ordered, one has to determine
whether the state contains such nonlocal correlations or entanglement. While
it is difficult to find the explicit form of the fattened string operators
$W_{\text{fat}}$,\cite{HWcnt} one can establish their existence or
non-existence using quantum information theory. The idea is that if the string
operators exist, then the entropy of an annular region (such as $A_1$ in Fig.
\ref{Z2topent1}) will be lower than one would expect based on local
correlations. 

The combination $(S_1 - S_2)-(S_3-S_4)$ measures exactly this anomalous
entropy. To see this, notice that $(S_1 - S_2)$ is the amount of additional
entropy associated with closing the region $A_2$ at the top.  Similarly,
$(S_3-S_4)$ is the amount of additional entropy associated with closing the
region $A_4$ at the top. If $\Psi$ has only local correlations with correlation
length $\xi$ then these two quantities are the same up to corrections of order
$O(e^{-R/\xi})$, since $A_2,A_4$ only differ by the region at the bottom. For
such states, $\lim_{R \rightarrow \infty} (S_1-S_2) - (S_3-S_4) = 0$. Thus, a
\emph{nonzero} value for $S_{\text{top}}$ signals the presence of nonlocal
correlations and topological order.

The universality of $S_{\text{top}}$ can also be understood from this picture.
Small deformations of $\Psi$ will typically modify the form of the string
operators $W_{\text{fat}}$ and change their width $l$. However, as long as $l$
remains finite, $(S_1-S_2)-(S_3-S_4)$ will converge to the same universal
number when the width $r$ of the annular region is larger than $l$.

\begin{figure}[tb]
\centerline{
\includegraphics[scale=0.4]{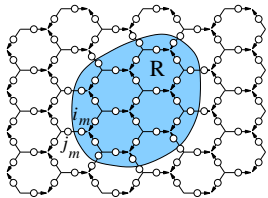}
}
\caption{
A simply connected region $R$ in the honeycomb lattice. 
We split the sites on the boundary links into two sites 
labeled $\v{i_m}$ and $\v{j_m}$, where $m=1,...,n$.
}
\label{smpspl}
\end{figure}

\textsl{A simple example}:
Let us compute the topological entropy of the ground state wave 
function $\Psi$ of the $Z_2$ model and confirm \Eq{quantdim} in this case.
We will first compute the entanglement entropy $S_{R}$ for an arbitrary region
$R$. To make the boundary more symmetric, we split the sites on the boundary 
links into two sites (see Fig. \ref{smpspl}). The wave function $\Psi$ 
generalizes to the new lattice in the natural way. The new wave function (still
denoted by $\Psi$) has the same entanglement entropy. 

We will decompose $\Psi$ into 
$\Psi = \sum_{l} \Psi_{l}^{\text{in}} \Psi_{l}^{\text{out}}$ where 
$\Psi_{l}^{\text{in}}$ are wave functions of spins inside $R$,  
$\Psi_{l}^{\text{out}}$ are wave functions of spins outside $R$, and $l$ is
a dummy index. A simple decomposition can be obtained using the string picture.
For any $q_{1},..., q_{n}$, with $q_{m} = 0,1$, and $\sum_m q_{m}$ even, we can
define a wave function $\Psi^{\text{in}}_{q_1,...,q_n}$ on the spins inside of 
$R$: $\Psi^{\text{in}}_{q_1,...,q_n}(X) = 1$ if (a) the strings in $X$ form 
closed loops and (b) $X$ satisfies the boundary condition that there is a 
string on $\v{i_m}$ if $q_m = 1$, and no string if $q_m = 0$. Similarly, we can
define a set of wave functions $\Psi^{\text{out}}_{r_1...,r_n}$ on the spins 
outside of $R$.

If we glue $\Psi^{in}$ and $\Psi^{out}$ together - setting $q_m = r_m$ for all
$m$ - the result is $\Psi$. Formally, this means that
\begin{equation}
\Psi = \sum_{q_1+...+q_l \text{ even}}\Psi^{\text{in}}_{q_1,...,q_n}
\Psi^{\text{out}}_{q_1...,q_n}
\label{decomp2}
\end{equation}
It is not hard to see that the functions $\{\Psi^{\text{in}}_{q_1,...,q_n} :
\sum_m q_m \text{ even}\}$, and $\{\Psi^{\text{out}}_{r_1...,r_n}: \sum_m r_m
\text{ even}\}$ are orthonormal up to an irrelevant normalization factor.
Therefore, the density matrix for the region $R$ is an equal weight mixture of
all the $\{\Psi^{\text{in}}_{q_1...,q_n}: \sum_m q_m \text{ even}\}$. There are
$2^{n-1}$ such states. The entropy is therefore $S_R = (n-1) \log 2$.
\cite{HIZ0515}

This formula applies to simply connected regions like the one in Fig.
\ref{smpspl}.  The same argument can be applied to general regions $R$ and
leads to $S_R = (n-j) \log 2$, where $n$ is the number of spins along $\partial
R$, and $j$ is the number of disconnected boundary curves in $\partial R$,

We are now ready to calculate the topological entropy associated with
$\Psi$. According to \Eq{quantdim} we need to calculate the entropy associated 
with the four regions shown in Fig. \ref{Z2topent1}. 
From $S_R = (n-j) \log 2$,
we find
$S_1 = (n_1-2) \log 2$,
$S_2 = (n_2-1) \log 2$, 
$S_3 = (n_3-1) \log 2$, and
$S_4 = (n_4-2) \log 2$,
where $n_1,n_2,n_3,n_4$ are the number of spins along the boundaries
of the four regions. The topological entropy is therefore
$-S_{\text{top}} = (n_1-n_2-n_3+n_4-2) \log 2$.
But the four regions are chosen such that $(n_1 - n_2) = (n_3 - n_4)$.  Thus
the size dependent factor cancels out and we are left with
$-S_{\text{top}} = -2 \log 2 = -\log(2^2)$.
This is in agreement with \Eq{quantdim} since the quantum dimension of 
$Z_2$ gauge theory is $D=2$. 

\textsl{General string-net models}:
To derive (\ref{quantdim}) in the general case, we compute the the topological 
entropy for the exactly soluble string-net models discussed in \Ref{LWstrnet}. 
The ground states of these models describe a large class of $(2+1)$ dimensional
topological orders. The  models and the associated topological orders are 
characterized by several pieces of data: (a) An integer $N$ - the number of 
string types. (b) A completely symmetric tensor $\del_{ijk}$ where 
$i,j,k = 0,1,...,N$ and $\del_{ijk}$ only takes on the values $0$ or $1$. This 
tensor represents the branching rules: three string types $i,j,k$ are allowed 
to meet at a point if and only if $\del_{ijk} = 1$. (c) A dual string type 
$i^*$ corresponding to each string type $i$. This dual string type corresponds 
to the same string, but with the opposite orientation. (d) A real tensor $d_i$ 
and a complex tensor $F^{ijm}_{kln}$ satisfying certain algebraic relations 
\cite{LWstrnet}. For each set of $F^{ijm}_{kln},d_i,\del_{ijk}$ satisfying 
these relations, there is a corresponding exactly soluble topologically ordered
spin model.

\begin{figure}[tb]
\centerline{
\includegraphics[width=1.6in]{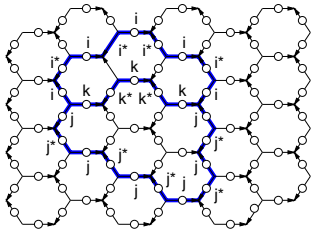}
}
\caption{
A typical string-net state on the honeycomb lattice. The empty links 
correspond to spins in the $i=0$ state. The orientation conventions on the
links are denoted by arrows.
}
\label{SSpin}
\end{figure}

The spins in the model are located on the links $\v{k}$ of the honeycomb
lattice. However, the spins are not usual spin-$1/2$ spins. Each spin can be in
$N+1$ different states which we will label by $i = 0,1,...,N$. Each spin state
can be thought of as a string-net state. To do this, one first needs to pick an
orientation for each link on the honeycomb lattice. When a spin is in state
$i$, we think of the link as being occupied by a type-$i$ string oriented in
the appropriate direction. If a spin is in state $i=0$, then we think of the
link as empty. In this way spin states correspond to string-net states (see
Fig. \ref{SSpin}). The Hamiltonian of the model involves a $12$ spin 
interaction \cite{LWstrnet}. The model is known to be gapped and topologically 
ordered and all the relevant quantities - ground state degeneracies, 
quasiparticle statistics, etc., can be calculated explicitly.

The ground state wave function $\Phi$ of the model is also known exactly. It is
easiest to describe in terms of the string-net language. If a spin 
configuration $\{i_{\v k}\}$ corresponds to an invalid string-net configuration
- that is, a string-net configurations that doesn't obey the branching rules 
defined by $\del_{ijk}$ - then $\Phi(\{i_{\v k}\}) = 0$. On the other hand, if 
$\{i_{\v k}\}$ corresponds to a valid string-net configuration then the 
amplitude is in general nonzero. We would like to have an explicit formula for 
these amplitudes. Unfortunately, this is not possible in general. However, we 
can write down linear relations that determine the amplitudes uniquely. These
relations relate the amplitudes of string-net configurations that only differ
by small local transformations. The relations are given by
\begin{align}
 \Phi
\bpm \includegraphics[height=0.3in]{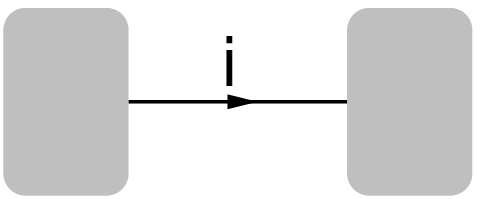} \epm  =&
\Phi 
\bpm \includegraphics[height=0.3in]{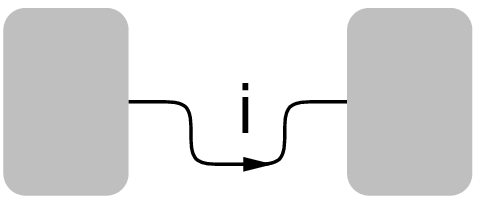} \epm
\label{topinv}
\\
 \Phi
\bpm \includegraphics[height=0.3in]{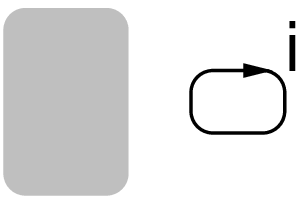} \epm  =&
d_i\Phi 
\bpm \includegraphics[height=0.3in]{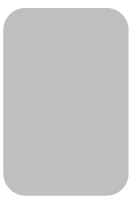} \epm
\label{clsdst}
\\
 \Phi
\bpm \includegraphics[height=0.3in]{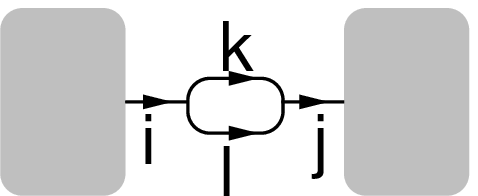} \epm  =&
\delta_{ij}
\Phi 
\bpm \includegraphics[height=0.3in]{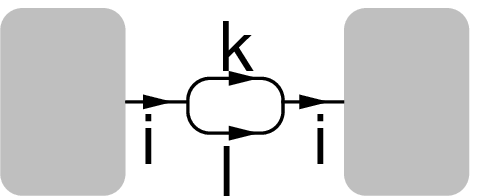} \epm
\label{bubble}\\
 \Phi
\bpm \includegraphics[height=0.3in]{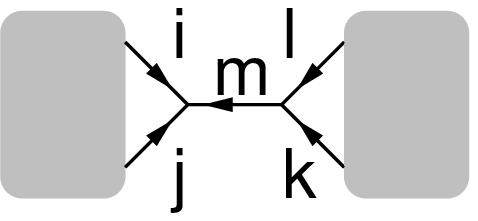} \epm  =&
\sum_{n} 
F^{ijm}_{kln}
\Phi 
\bpm \includegraphics[height=0.3in]{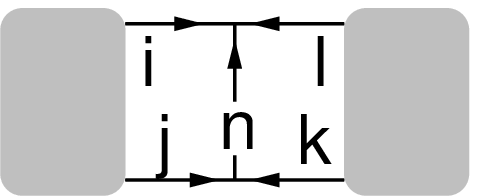} \epm
\label{fusion} 
\end{align}
where the shaded areas represent other parts of the string-nets that are not
changed. Also, the type-$0$ string is interpreted as the no-string (or vacuum)
state. The first relation (\ref{topinv}) is drawn schematically. The more 
precise statement of this rule is that any two string-net configurations 
on the honeycomb lattice that can be continuously deformed into each other have
the same amplitude. In other words, the string-net wave function $\Phi$ only 
depends on the topologies of the network of strings.

By applying these relations multiple times, one can compute the amplitude for
any string-net configuration (on the honeycomb lattice) in terms of the 
amplitude of the vacuum configuration. Thus, (\ref{topinv}-\ref{fusion}) 
completely specify the ground state wave function $\Phi$. 

Let us first compute the von Neumann entropy $S_R$ of the exact ground state
wave function $\Phi$ for a simply connected region $R$ (see Fig.
\ref{smpspl}). Again we split the site on the boundary links into two
sites.  We decompose $\Phi$ into $\Phi = \sum_{l} \Phi_{l}^{\text{in}}
\Phi_{l}^{\text{out}}$ where $\Phi_{l}^{\text{in}}$ are wave functions of spins
inside $R$,  $\Phi_{l}^{\text{out}}$ are wave functions of spins outside $R$,
and $l$ is some dummy index. 

A wave function $\Phi^{\text{in}}$ on the spins inside of $R$ can be defined as
follows. Let $\{i_{\v k}\}$ be some spin configuration inside of $R$. If 
$\{i_{\v k}\}$ doesn't correspond to a valid string-net configuration - that is
one that obeys the branching rules, then we define 
$\Phi^{\text{in}}(\{i_{\v k}\})=0$. If $\{i_{\v k}\}$ does correspond to a 
valid string-net configuration, then we define $\Phi^{\text{in}}(\{i_{\v k}\})$
using the same graphical rules (\ref{topinv}-\ref{fusion}) that we used 
for $\Phi$. 

However, there is an additional subtlety. Recall that in the case of $\Phi$, 
the graphical rules could be used to reduce any string-net configuration to the
vacuum configuration. To fix $\Phi$, we defined $\Phi(\text{vacuum}) = 1$. 

\begin{figure}[tb]
\centerline{
\includegraphics[scale=0.35]{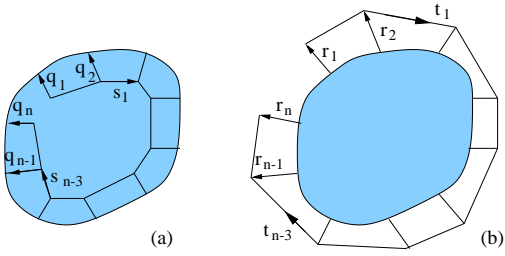}
}
\caption{
The basic string-net configurations (a) $X_{\{q,s\}}$ for inside $R$ and 
(b) $Y_{\{r,t\}}$ for outside $R$. 
}
\label{inandout}
\end{figure}

In this case, since we are dealing with a region $R$ \emph{with a boundary}, 
string-net configurations cannot generally be reduced to the vacuum 
configuration. However, they can be reduced to the 
tree-like diagrams $X_{\{q,s\}}$ shown in Fig. \ref{inandout}a. Thus, to
define $\Phi^\text{in}$, we need to specify the amplitude for all of these 
basic configurations. There are multiple ways of doing this and hence multiple
possibilities for $\Phi^\text{in}$. Here, we will consider all the 
possibilities. For any labeling $q_1,...,q_n,s_1,...,s_{n-3}$ of the string-net
in Fig. \ref{inandout}(a), we define a wave function 
$\Phi^{\text{in}}_{\{q,s\}}$ by
$\Phi^{\text{in}}_{\{q,s\}}(X_{\{q',s'\}}) = \del_{\{q\},\{q'\}}
\del_{\{s\},\{s'\}}$.
Starting from these amplitudes and using the graphical rules 
(\ref{topinv}-\ref{fusion}) we can determine $\Phi^{\text{in}}_{\{q,s\}}(X)$ 
for all other string-net configurations. In the same way, we can define wave 
functions $\Phi^{\text{out}}_{\{r,t\}}$ on the spins outside of $R$ through
$\Phi^{\text{out}}_{\{r,t\}}(Y_{\{r',t'\}}) = \del_{\{r\},\{r'\}}
\del_{\{t\},\{t'\}}$, where the $Y_{\{r,t\}}$ are shown in 
Fig. \ref{inandout}(b).

\begin{figure}[tb]
\centerline{
\includegraphics[scale=0.35]{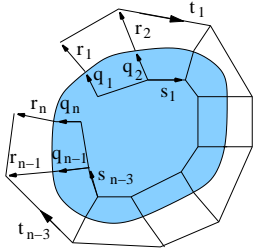}
}
\caption{
The string-net configuration $Z_{\{q,s,r,t\}}$ obtained by ``gluing''
the configuration $X_{\{q,s\}}$ to the configuration
$Y_{\{r,t\}}$ in Fig. \ref{inandout}.
}
\label{inoutreg}
\end{figure}

Now consider the product wave functions
$\Phi^{\text{in}}_{\{q,s\}}\Phi^{\text{out}}_{\{r,t\}}$.  These are wave
functions on the all the spins in the system - both inside and outside $R$.
They can be generated from the amplitudes for the string-net
configurations $Z_{\{q,s,r,t\}}$ in Fig. \ref{inoutreg}:
\begin{equation*}
\Phi^{\text{in}}_{\{q,s\}}\Phi^{\text{out}}_{\{r,t\}}
(Z_{\{q',s',r',t'\}})
= 
\del_{\{q\},\{q'\}}\del_{\{s\},\{s'\}}
\del_{\{r\},\{r'\}}\del_{\{t\},\{t'\}}
\end{equation*}

On the other hand, it is not hard to show that for the ground state wave 
function $\Phi$, the amplitude for $Z_{\{q,s,r,t\}}$ is
\begin{displaymath}
\Phi(Z_{\{q,s,r,t\}}) = 
\del_{\{q\},\{r\}} \del_{\{s\},\{t\}} \prod_{m} (\sqrt{d_{q_m}})
\end{displaymath}
Comparing the two, we see that
\begin{equation}
\Phi = \sum_{\{q,s,r,t\},}
\Phi^{\text{in}}_{\{q,s\}}\Phi^{\text{out}}_{\{r,t\}} 
\del_{\{q\},\{r\}} \del_{\{s\},\{t\}}
\prod_m (\sqrt{d_{q_m}})
\end{equation}

It turns out that the wave functions $\{\Phi^{\text{in}}_{\{q,s\}}\}$
are orthonormal, as are the $\{\Phi^{\text{out}}_{\{r,t\}}\}$ (up to an 
irrelevant normalization constant). This means that we can use them as a basis.
If we denote $\Phi^{\text{in}}_{\{q,s\}}\Phi^{\text{out}}_{\{r,t\}}$
by $|\{q,s,r,t\}\>$, then in this basis, the wave function $\Phi$ is
\begin{align}
\<\{q,s,r,t\}|\Phi\> = 
\del_{\{q\},\{r\}} \del_{\{s\},\{t\}}
\prod_m (\sqrt{d_{q_m}})
\end{align}
The density matrix for the region $R$ can now be obtained by tracing out the
spins outside of $R$, or equivalently, tracing out the spin states
$|\{r,t\}\>$:
\begin{equation}
\<\{q',s'\}|\rho_R|\{q,s\}\> = 
\del_{\{q\},\{q'\}} \del_{\{s\},\{s'\}} \prod_m d_{q_m}
\end{equation}
Since the density matrix is diagonal, we can easily obtain the entanglement
entropy for $S_R$. Normalizing $\rho_R$ so that $\Tr(\rho_R) =1$, and
taking $-\Tr \rho_R \log \rho_R$, we find
\begin{equation}
S_R =  -\sum_{\{q,s\}} \frac{\prod_{m} d_{q_m}}{D^{n-1}}
\log \left(\frac{\prod_{l} d_{q_l}}{D^{n-1}}\right) \\
\end{equation}
where $D = \sum_k d_k^2$. The sum can be evaluated explicitly (with the help of
the relations in \cite{LWstrnet}). The result is
\begin{equation}
S_R = -\log(D) - n \sum_{k=0}^N \frac{d_k^2}{D} \log\left(\frac{d_k}{D} \right)
\end{equation}
This result applies to simply connected regions like the one shown in Fig.
\ref{Z2topent1}. The same argument can be applied to general regions $R$. In 
the general case, we find
\begin{equation}
S_R = -j\log(D) - n\sum_{k=0}^N \frac{d_k^2}{D} \log\left(\frac{d_k}{D} \right)
\label{strnetent}
\end{equation}
where $n$ is the number of spins along $\partial R$, and $j$ is the
number of disconnected boundary curves in $\partial R$.

We can now calculate the topological entropy associated with $\Phi$.  
Applying (\ref{strnetent}), we find $S_1 = -2\log D - n_1 s_0$, 
$S_2 = -\log D  - n_2 s_0$, $S_3 = -\log D  - n_3 s_0$, and 
$S_4 = -2\log D - n_4 s_0$ where $n_1,n_2,n_3,n_4$ are the numbers of spins 
along the boundaries of the four regions, and 
$s_0 = \sum_{k=0}^N \frac{d_k^2}{D} \log\left(\frac{d_k}{D} \right)$. 
The topological entropy is therefore
$-S_{\text{top}} = -2\log D + (n_1-n_2-n_3+n_4) s_0 = -2\log D$
in agreement with \Eq{quantdim}.

Near the completion of this paper, we become aware of a similar result,
obtained independently in the recent paper, \Ref{KP0592}.
This research is supported by NSF Grant No. DMR--04--33632 and by 
ARO Grant No. W911NF-05-1-0474.


\end{document}